\documentclass[runningheads]{llncs}

\usepackage{times}
\usepackage{epsfig}
\usepackage{graphicx}
\usepackage{amsmath}
\usepackage{amssymb}
\usepackage[utf8]{inputenc}
\pdfoutput=1
\usepackage{dblfloatfix}
\usepackage{float}

\usepackage{fourier}

\usepackage[pagebackref=true,breaklinks=true,colorlinks,bookmarks=false]{hyperref}

\begin{document}

\title{Climbing Routes Clustering Using Energy-Efficient Accelerometers Attached to the Quickdraws}

\author{Sadaf Moaveninejad \inst{1}\orcidID{0000-0003-2144-4412} \and
Andrea Janes\inst{2}\orcidID{0000-0002-1423-6773} \and
Camillo Porcaro\inst{1}\orcidID{0000-0003-4847-163X}\and
Luca Barletta\inst{3}\orcidID{0000-0003-4052-2092}\and
Lorenzo Mucchi\inst{4}\orcidID{0000-0001-6389-0221}\and
Massimiliano Pierobon\inst{5}\orcidID{0000-0003-1074-6925}
}

\authorrunning{S. Moaveninejad et al.}
\institute{Department of Neuroscience and Padova Neuroscience Center, University of Padova, Padova, Italy\\
\email{\{sadaf.moaveninejad, camillo.porcaro\}@unipd.it}
\and
FHV Vorarlberg University of Applied Sciences, Dornbirn, Austria\\
\email{andrea.janes@fhv.at}
\and
Department of Electronics, Information and Bioengineering, Politecnico di Milano, Milan, Italy\\
\email{luca.barletta@polimi.it}
\and
Department of Information Engineering, University of Florence, Firenze, Italy\\
\email{lorenzo.mucchi@unifi.it}
\and
School of Computing, University of Nebraska-Lincoln, Lincoln, Nebraska, USA
\email{maxp@unl.edu}\\
}
\maketitle

\begin{abstract}
   One of the challenges for climbing gyms is to find out popular routes for the climbers to improve their services and optimally use their infrastructure. This problem must be addressed preserving both the privacy and convenience of the climbers and the costs of the gyms. To this aim, a hardware prototype is developed to collect data using accelerometer sensors attached to a piece of climbing equipment mounted on the wall, called quickdraw, that connects the climbing rope to the bolt anchors. The corresponding sensors are configured to be energy-efficient, hence becoming practical in terms of expenses and time consumption for replacement when used in large quantities in a climbing gym. This paper describes hardware specifications, studies data measured by the sensors in ultra-low power mode, detect patterns in data during climbing different routes, and develops an unsupervised approach for route clustering.
\end{abstract}
\vspace{-0.2cm}
\vspace{-0.1cm}
\section{Introduction}
In the past two decades, sport climbing has become a highly popular sport \cite{olympics}. Sport climbing refers to a form of climbing in which climbers use routes with prepared anchor points that prevent a climber from a ground fall. In this way, the challenge lies more in the mastery of the climbing route and less in maintaining safety. 

In this paper, when talking about sport climbing, we refer to lead climbing, i.e., a form of sport climbing in which climbers are not secured by a rope hanging from the top, but in which climbers  use the anchor points to attach the rope as they proceed to subsequent anchor points; one end of the rope is held by the belayer, the other end is attached at the climber.

In case of a fall, the rope is deflected on the last anchor point and the fall height of the climber minimized. On completion, climbers use the last (top) anchor point to secure the rope so that their partners on the ground can lower them. It can be carried out either outdoor on natural cliffs or indoor in climbing gyms. In the case of indoor, climbing gyms set up artificial walls with several routes with various difficulties. Along these routes, climbing gyms provide climbers with equipment and services which gives them the opportunity to challenge themselves and improve their skills without scarifying their safety. To this aim and similar to other sports (\textit{i.e.} cycling, running), engineering and science have begun to help sport climbing \cite{international2006engineering}. 

The contribution of science in sports activities could take place in the form of collecting data from athletes using sensors embedded in electronic equipment such as smartwatches, smart bands, fitness trackers, and smartphones. These devices are attached to the body of the athletes (as in \cite{Boulanger2016,Ladha2013,Kosmalla2015,Pansiot2008,boulanger2015automatic} ). The corresponding measurements are time-series data used to visualize statistics of athletes and analyze their performance. Such analysis could be done either by a coach or through certain applications. Another way of acquiring data to analyze sports activities is based on the camera. In this approach, computer vision algorithms are developed to extract human 2-dimensional (2D) pose sequences from video frames and based on skeleton estimation of each person \cite{einfalt2019frame,sasaki2020exemposer}. Instrumented climbing walls is the other approach for monitoring climbers which is inspected in \cite{aladdin2012static,fuss2008instrumented}

The data acquisition methods based on wearable-sensors and camera are not always desirable for climbers. The former limits the convenience of the user when wearing an extra device during climbing, and the latter is against privacy. Keeping these issues in mind, for this paper data is collected from accelerometer sensors attached to a piece of climbing equipment mounted on the wall, called quickdraw. The sensor enhanced quickdraw hereinafter is referred to as smart-quickdraw (s-qd). To the best knowledge of the authors, the sensor enhanced quickdraws were studied the first time in our previous work \cite{ivanova2020video}. We developed a hybrid system based on sensor and camera to detect rope-pulling activity. Apart from climbers, another concern for data collection relates to the climbing gyms. There is a large number of quickdraws in a climbing gym and regular changing batteries of the s-qds is expensive for the gym in terms of both time and costs. Hence, sensors attached to the quickdraws must be energy efficient such that their batteries do not need to be replaced in the short term. 

The data collected from climbers helps both climbers and gyms in different ways. One goal is to improve the performance of the climbers, to achieve this, first must detect different activities during climbing \textit{i.e.} ascending, resting, falling, lowering, and rope-pulling at the end of the climb. Then, the detected activities could be analyzed by experts and in off-line modes to provide recommendations for better performance, or in on-line mode to improve safety by early prediction of risk. Another goal is to improve the infrastructure of the gyms based on climbers' needs. 

In this work, we investigated data from s-qds on the wall to differentiate between the different routes climbed in the same line. This reveals the most popular routes and helps the climbing gyms to better understand the needs of the climbers. Authors in \cite{Kosmalla2015} introduced a system to automatically detect the routes based on arm orientation from a single sensor. They collected data from wrist-worn Inertia Measurement Units (IMUs), extracted features from the sensor measurements, and finally evaluate the method by cross-validation. Our proposed method is an unsupervised algorithm based on clustering which does not need prior training of the model. This property makes our algorithm more general and less complex for the user since it could be easily adapted for new lines with different routes. 

However, using sensors in energy-efficient mode and attaching them to the wall, instead of the body of the climber, introduces some challenges. Saving energy reduces the number of sensor samples. Moreover, different from algorithms based on wearable devices which mainly collect data through a single sensor on the body of the climber, we need to deal with multiple sensors when they are on the wall. Additionally, each sensor on the wall can provide us useful data only in a certain period during the climb.
In the following, all these points are addressed in detail.

\subsection{Contributions}
\label{subsec:Contributions}
Two male climbers participated in the experiment. The participants' skill levels were advanced i.e., self-estimated as a and 6b on-sight on the French Rating Scale of Difficulty (FRSD). The FRSD is a widely recognized system in the climbing community for grading the difficulty of routes. It ranges from easy levels (such as 1 or 2) to extremely challenging levels (like 9a). The scale progresses linearly: as the number increases, so does the difficulty. For example, a 6b route is more challenging than a 5c.
A route graded 6b+ is a step above 6b, presenting even more complex and physically demanding challenges.

In our experiment, we selected routes graded 5c+, 6a+, and 6b+ to match the advanced skill levels of our participants. These grades ensured that the routes were challenging enough to test their abilities, yet within their competency range.

One participant was 27 years old with ten years of climbing experience and the other of 65 years old climbing for thirty years. For the purpose of data collection, the participants were asked to climb in the leading style on three different days. The three pre-selected routes were with difficulty levels of 5c+,6a+,and 6b+ (in French grading system). The routes were selected according to the skill levels of both climbers. The participants climbed at their usual speed, clipping the rope into every quickdraw, and were free to take resting time between climbs. The participant who climbed pulled the rope after each ascent. There was a person responsible for belaying. 

On the first day, climber 1 was asked to climb a line with three routes nine times (each route three times). The order of climbing was: (5c+, 5c+, 5c+), (6b+, 6b+, 6b+), and (6a+, 6a+, 6a+). The following days, climber 1 practiced more and after one week he was asked to climb again the same line. In this experiment, which we refer to as $2^{nd}$-day of climber1, he climbed each route five times. This time we changed the order of climbing as (5c+, 6a+, 6b+) and repeated this process five times. Finally on the $3^{rd}$-day, we asked climber 2 to repeat a similar experiment as the $1^{st}$-day of climber1 but with different order as (5c+, 5c+, 5c+), (6a+, 6a+, 6a+), and (6b+, 6b+, 6b+). The aim of changing the climbers and the order of routes was to add different parameters \textit{i.e} age, tiredness, and provisioning to the data collected from the same routes.

Employing sensors in ultra-low-power mode enhances energy saving, but instead significantly reduces the number of samples. Considering this fact, we need to carefully understand the sensor functionality to analyze the received samples and find the more informative ones for our objective. To this aim, we described the details of our data acquisition system in Sec.~\ref{Sec:Data Acquisition System}. Afterward, in Sec.~\ref{Sec:Data Analysis}, the relevant features for route detection are extracted from sensor measurements. However, not all the features are necessary and a subset of features is selected through features optimization as in Sec.~\ref{Sec:Features Optimization}. In Sec.~\ref{Sec:Routes clustering} the selected features are used for clustering the climbs to $3$ different routes and the corresponding results are discussed in Sec.~\ref{Sec:Features Optimization}. Finally, the conclusion and future works are discussed in Sec.~\ref{Sec:Conclusion}.
\section{Data Acquisition System}
\label{Sec:Data Acquisition System}
\begin{figure}[h]
    \centering
    \vspace{-0.3cm}
    \includegraphics[width=0.7\textwidth]{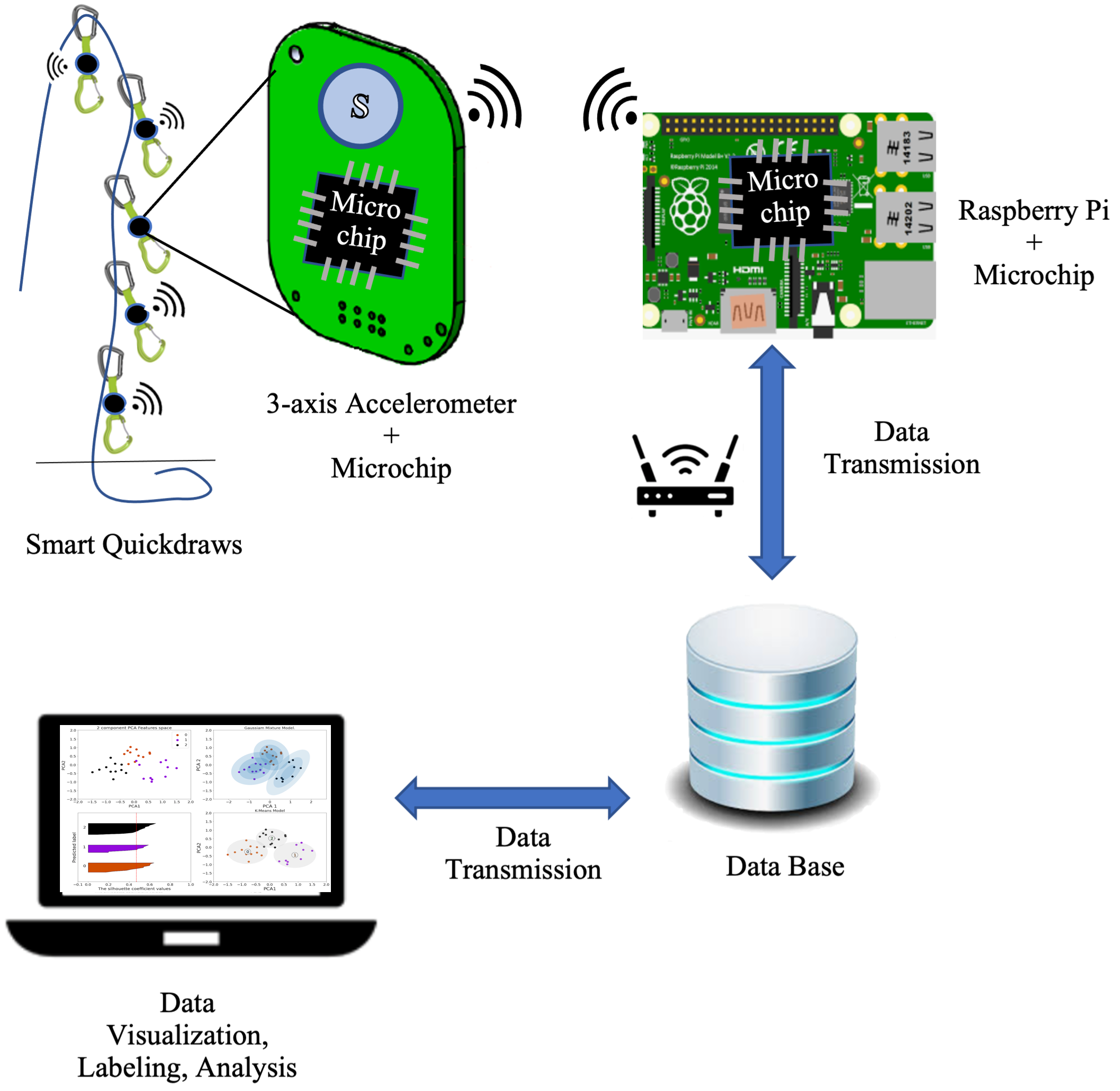}
    \vspace{-0.2cm}
    \caption{Data acquisition system}
    \vspace{-0.5cm}
    \label{fig:iot}
\end{figure}

Our data acquisition system is shown in Fig \ref{fig:iot} and consists of two main parts: 1- smart-quickdraws to measure 3-axis accelerations, 2- a base station to receive data from sensors and forwards them to the database in a remote server.
\subsection{Smart-quickdraw}
The so-called smart-quickdraw essentially refers to an in-house circuit board that is attached to the strip in the central part of the quickdraw. This board consists of an 3-axial accelerometer sensor to capture quickdraw movements along x, y, and z directions as shown in Fig \ref{fig:sensor_xyz} and a microchip to control the accelerometer and communicate with the base station Fig \ref{fig:iot}. For our application, energy efficiency is a primary issue as battery replacement of all smart-quickdraws in a climbing gym is both time and cost-consuming. In this regard, for the accelerometer, we selected LIS3DH by STMicroelectronics \cite{lis3dh} which could be configured to operate in ultra-low-power modes through smart sleep-to-wake-up and return-to-sleep functions. Concerning the microchip, we used ATSAMR21G18A by Atmel \cite{atsamr21g18a_microchip} which combines a microcontroller unit (MCU) and an RF transceiver. The accelerometer sensor is controlled and configured via firmware which is developed inside the MCU, moreover, the smart-quickdraw communicates with the base station via the transceiver. In the following, some of our main settings for the accelerometer sensors are listed: 
\begin{figure}[t!]
    \centering
    \vspace{-0.3cm}
    \includegraphics[width=0.25\textwidth]{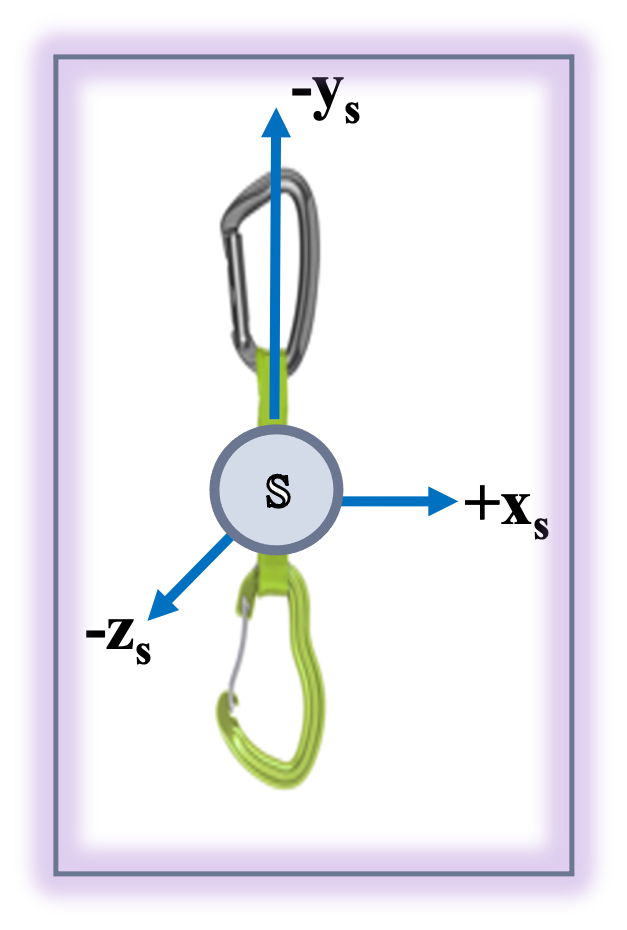}
    \vspace{-0.2cm}
    \caption{x,y, z directions of 3-axis accelerometer sensor with respect to the climbing wall}
    \vspace{-0.2cm}
    \label{fig:sensor_xyz}
\end{figure}
\begin{itemize}
    \item {\bf Full scale:} 
    The LIS3DH has a dynamic user-selectable range of forces it can measure which are from ±2g to ±16g. Typically in accelerometers, the smaller the range, the more sensitive the readings will be. Hence, we selected 2g as the full scale for our prototype.
    \item {\bf Data rate:}
    The LIS3DH is capable of measuring accelerations with output data rates from 1 Hz to 5 kHz. For our energy-efficient system, we selected 10 Hz and 50 Hz for the inactive and active modes, respectively.
    \item {\bf Output bits:} 
    The analog to digital converter (ADC) of LIS3DH could have 8, 10, or 12 bits data output for operating in low-power, normal and high-resolution modes. Since our goal is to have an energy-efficient system, we selected 8-bit resolution for the sensor data output.
    \item {\bf Range and resolution:}
    Accelerometers usually provide raw values which are not equivalent to meters per second squared. Consequently, we still need to scale the accelerometers' output based on our setting for full scale. In this sensor's case, a signed 8-bit output data corresponds to a number ranging from -127 and 127. After scaling this range by ±2g full scale, sensor data output 127 is +2g of force, and -127 is -2g. The resolution of the data output after scaling is $\frac{2g}{127} \approx 16 mg$. Hence, knowing the range and scale is a key to deciphering the sensor's data output.
    \item {\bf Noise reduction:}
    At a sampling rate of 50 Hz, there is still considerable noise in the sensor readings, which is the reason why a software averaging method has been implemented in our prototype. This simple method takes every subsequent set of 8 samples and calculates the average for the acceleration on each axis. Then, the corresponding averaged value is transmitted to the raspberry-pi.
    \item {\bf Filter insignificant changes:}
    Regardless of whether there is movement or not, the accelerometer always provides sensor readings for the configured sampling rate. Since bandwidth is a premium for low-power devices, sensor values without any changes on any axis are not transmitted to the base station. However, even after averaging, hardly any sample is the same except if the sensor is completely still. Therefore a method has been implemented that prevents samples to be sent if there is no absolute change on any axis below a certain threshold. For this prototype, we set a threshold value equal to 15 units of the sensor's data output. The sensor does not consider any sample worth transmitting if the measured value of none of the axis has changed by at least 15 units. As mentioned above, the resolution of the data output for this sensor configuration is almost 16mg, therefore the 15 units threshold is equivalent to the acceleration of $15 \times 16\ mg = 240 \ mg$. This value is chosen because $240 \ mg$ represents a significant enough movement to be relevant for analysis, while also being large enough to filter out minor fluctuations that could be attributed to noise.
    \item {\bf Grouping samples:}
    To save even more power (the radio of the smart-quickdraw is always off while no data is being sent) all sensor values are sent to the base station as a batch of values, i.e., grouped. We set this value to two samples, meaning that no sample is sent out individually, except before going to sleep after inactivity, in that case, any sample that was held back is being sent out. This may have the effect that sensor data is not received to the base station in real-time, but with a delay. However, since every sensor value has a timestamp this should not be an issue for analysis use-cases.
\end{itemize}
\subsection{Base station}
The base station of our system comprises a board with the microchip, similar as in the smart-quickdraw, mounted on a raspberry-pi \cite{raspberrypi}. The raspberry-pi is used as a powerful CPU and hosts the application programming interface (API) that allows communication with the devices, as well as a web-server and possibly other services. In addition, the raspberry-pi CPU communicates with its board via the serial line. The firmware of the board within the base station performs the "network translation" to the radio network that is used amongst the boards (inside the base station and smart-quickdraws).

\subsection{Ultra-low-power data acquisition:}
\label{subsec:Ultra-low-power data acquisition}
In our system, the power-saving features are primarily managed by the CPU within each smart-quickdraw, which is a part of the Microcontroller Unit. This CPU is responsible for configuring the sensor's operation modes between sleep and active modes and is distinct from the CPU in the base station, which has a different role and does not interact directly with the sensor or its configuration settings. The s-qd CPU functions as follows:
\begin{itemize}
    \item {\bf Sleep mode:} When the sensor is stationary, the CPU sets the sensor to sleep mode, sampling data at a lower rate (10 Hz). This conserves power as less data is processed and transmitted.
    \item {\bf Activation and inactivity detection:} If the sensor detects movement exceeding the threshold (15 units of sensor data), the CPU switches the sensor to active mode (50 Hz sampling rate) for more frequent data collection.
    \item {\bf Inactivity Detection:} 
    \begin{enumerate}
        \item If the sensor data exceeds the threshold (15 units of sensor data), indicating significant movement, the CPU switches the sensor to active mode, increasing the sampling rate to 50 Hz for more frequent data collection.
        \item Conversely, if the measured values in all three axes are consistently lower than the threshold, the CPU filters out these readings as insignificant, preventing unnecessary data processing.
    \end{enumerate}
    \item {\bf Transition Back to Sleep Mode: } 
    While in active mode, the CPU averages every 8 samples and sends them to the base station. If the averaged data falls below the threshold for 0.8 seconds, the sensor is considered inactive. If this inactivity continues for 20 seconds, the CPU switches the sensor back to sleep mode.
\end{itemize}
In contrast, the base station CPU primarily manages data reception and network communication, without influencing the sensor's power-saving configurations.


\section{Features Engineering}
\label{Sec:Data Analysis}
After visually keeping track of the climbers in the gym, we assume that the most relevant information for route identification are: 1- statistical features of quickdraws accelerations, 2- the moment when the climber reaches each quickdraw. Accordingly, we extract statistical and temporal features from sensor measurements as stated in the following. 
\subsection{Statistical Features}
The sensor embedded in smart-quickdraw works in ultra-low-power mode, hence the s-qd does not continuously send data to the base station. Instead, transmits samples when it is in active mode and the change in the movement of the corresponding quickdraw exceeds a certain threshold. In our experience, all quickdraws of the line are sensor enhanced, and $i, j$ refer to the position of two different s-qds in the line. During climb $c$, each sensor at position $i$ transmits samples with timestamp $t_{{i}_{k}}$:
\begin{equation}
{T_{i}}^{c} = \left \{ t_{{i}_{k}} | k\geqslant 1 \right \}
\end{equation}
where $k$ refers to the index of the samples and ${T_{i}}^{c}$ is the set of timestamps of all samples transmitted from s-qd at position $i$ to the base station. These samples are measured when the s-qd is clipped till shortly afterward, while the climber's movement still causes tangible changes in the acceleration of the s-qd. On this assumptions, a vector of statistical features could be extracted from the $N_{i}^{c}$ samples measured by each s-qd in a period between the moment when s-qd is activated through clipping at $t_{i_{k=1}}$, till clipping the next sensor at $t_{(i+1)_{k=1}}$. In addition to the 3-axis accelerations along the x,y, z axis, for each sample we can have a new value $g$ corresponds to the sum of the energy of that sample along the three axes:
\begin{equation}
    g_{i_k}^{c} = \sqrt{({x_{i_k}^{c}})^{2}+({y_{i_k}^{c}})^{2}+({z_{i_k}^{c}})^{2}}\, , \;\;  1\leq  k < N_{i}^{c}
\end{equation}
Accordingly, we have four data sets for each s-qd:
\begin{align}
    X_{i}^{c} & = \left \{ x_{i_k}^{c} \, | \, 1\leq  k < N_{i}^{c} \right \} \\
    Y_{i}^{c} & = \left \{ y_{i_k}^{c} \, | \, 1\leq  k < N_{i}^{c} \right \} \\
    Z_{i}^{c} & = \left \{ z_{i_k}^{c} \, | \, 1\leq  k < N_{i}^{c} \right \} \\
    G_{i}^{c} & = \left \{ g_{i_k}^{c} \, | \, 1\leq  k < N_{i}^{c} \right \} 
\end{align}
For each s-qd at position $i$, the following statistical features are calculated over $X_{i}^{c}$, $Y_{i}^{c}$, $Z_{i}^{c}$, and $G_{i}^{c}$: 1-mean, 2- minimum, 3- maximum, 4- variance, 5- standard deviation, 6- root mean square, 7- the $p^{th}$ percentile where $p = \left \{ 5, 25, 75, 95 \right \}$, 8- kurtosis, 9- skew, 10- Pearson correlation (between $x_{i_k}^{c}y_{i_k}^{c}$, $x_{i_k}^{c}z_{i_k}^{c}$, and $y_{i_k}^{c}z_{i_k}^{c}$), 11- number of peaks.

\subsection{Temporal Features}
Knowing the moment of clipping the rope to each s-qd, a vector of temporal features could be extracted from the timestamps of the received samples as follows:
\begin{enumerate}
    \item ${\Delta ts}^{c}$: a set of numbers which represents the time the climber spends in short-segments of the route between every two subsequent s-qds at positions $i$ and $i+1$:
    \begin{equation}
    \label{eq:delta_t_short}
        \Delta ts^{c} = \left \{ \Delta t_{i_{k},i+1_{k}}^{c} | i= 2,\cdots,(ie-2), k=1 \right \}
     \end{equation}
     where $ie$ is the position of the s-qd at the end of the line.
    \item ${\Delta tl}^{c}$: a set of numbers which represents the time a climber spends between starting point of the climb $i = 2$ and each s-qd $j $. These time-deltas correspond to long-segments of the route and excludes $j=i+1$ which is already considered in (\ref{eq:delta_t_short}):
    \begin{equation}
        \Delta tl^{c} = \left \{\Delta t_{i_{k},j_{k}}^{c} | i=2 , j=(i+2),\cdots,(ie-2), k=1 \right \}
        \label{eq:delta_t_long}
    \end{equation}
    \item ${\Delta tc}^{c}$: the duration of the climb $c$ which is equivalent to the time difference between clipping the rope to the s-qds in the beginning and at the end of the climb. This is the extreme case of (\ref{eq:delta_t_long}) where $j=ie-1$.
    \begin{equation}
        \Delta tc^{c} = \left \{\Delta t_{i_{k},j_{k}}^{c} | i=2 , j= ie-1, k=1 \right \}
        \label{eq:delta_t_climb}
    \end{equation}
    \item Statistical features of $\Delta {ts}^{c}$: a set of features including minimum, maximum, mean, and standard deviation of ${\Delta ts}^{c}$ (the time a climber spends in short segments of a route in each climb).
\end{enumerate}
\section{Features Optimization}
\label{Sec:Features Optimization}
In the previous section, we introduced the features which are intended to be used for route clustering. Before clustering, features are optimized in two steps: 1- feature scaling, 2- feature selection. 
\subsection{Features scaling}
Many machine learning algorithms (\textit{i.e.} Linear Regression, Logistic Regression, K-Means clustering) which calculate the similarities based on Euclidean distance do not give a reasonable recognition to the smaller feature \cite{saha2017feature}. Accordingly, features must be pre-processed through scaling to normalize them within a particular range. For instance, if not scaled, the time-delta of longer segments ${\Delta tl}^{c}$ dominate the ones corresponding to shorter segments ${\Delta ts}^{c}$. 

There are different scaling techniques, the most common ones are Standards scaling, Min-Max scaling, Robust scaling, and Quantile transform. We selected Quantile Transform scaler since this scaler converts the variable distribution to a normal one and spread out the most frequent values and reduces the impact of outliers \cite{scikit-learn}. 
\subsection{Feature selection}
\label{subsec:Feature selection}
\begin{figure}[t!]
    \vspace{-0.3cm}
    \includegraphics[width=0.85\textwidth]{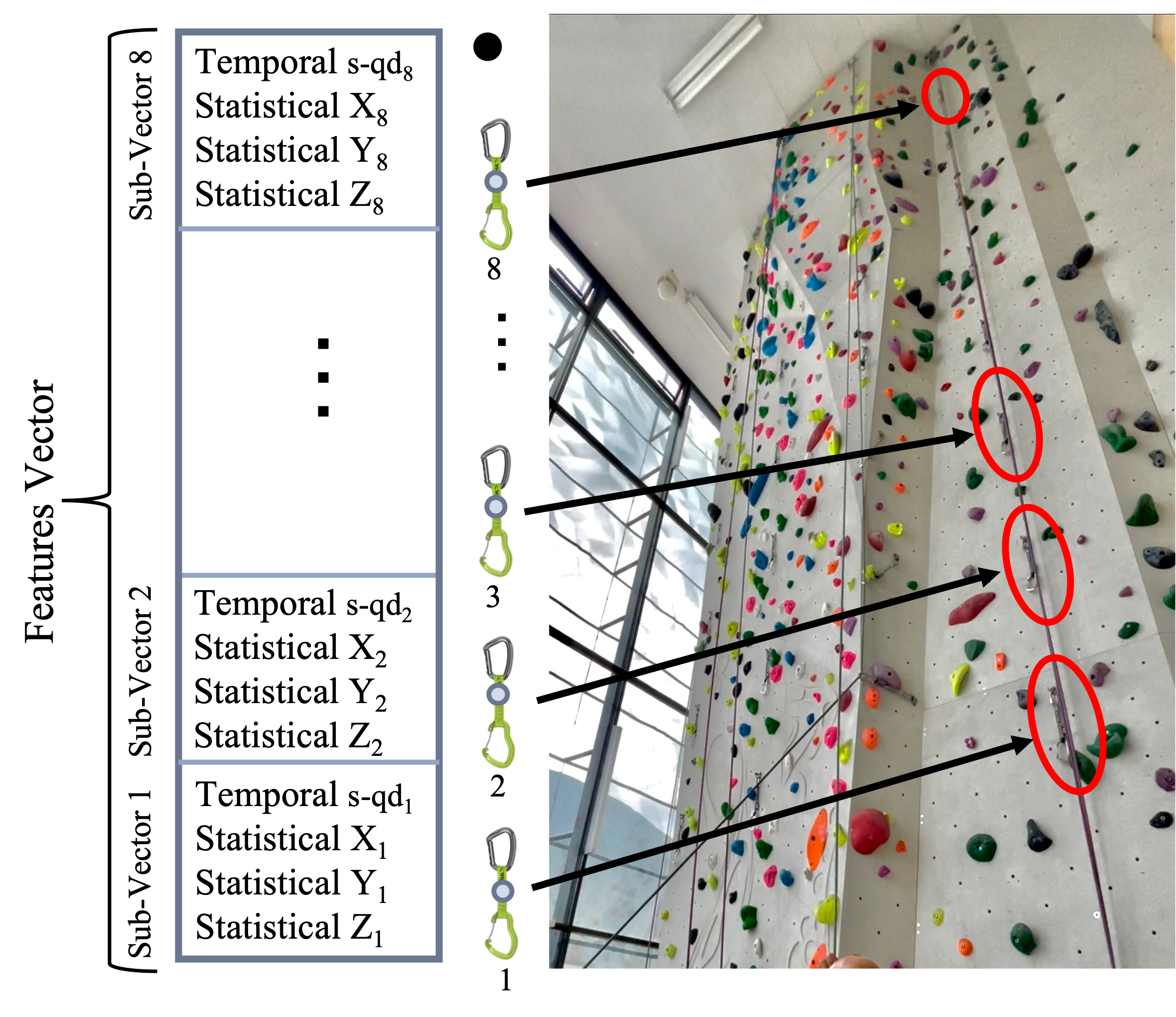}
    \vspace{-0.2cm}
    \caption{Features vector of one climb composed of sub-vectors of temporal and statistical features of each s-qds}
    \vspace{-0.2cm}
    \label{fig:features_vector}
\end{figure}

The length of the features vector depends on the number of s-qds. As shown in the Fig \ref{fig:features_vector}, features vector of each climb consists of sub-vectors extracted from s-qds. We ignored the sub-vectors related to the first and last s-qds. Because, the first sensor is mainly affected by the performance of the belayer. In addition, the last sensor measures the movements of the last s-qd when the climber finished climbing and is preparing for lowering. Consequently, our features vector consists of sub-vectors 2 to 7. However, some of these features (and corresponding s-qds) are redundant because different routes in the same line differ in certain parts and have similar difficulties in the rest of the line. In this regard, the features vector must be optimized not only to reduce the number of features and increase the efficiency and effectiveness of the algorithm but also to simplify the hardware by removing sensors that provide irrelevant features. 
In this work, we utilized univariate feature selection to find features that can cover data related to different climbers, or the same climber on different days. This approach works by assigning a score to all features, based on univariate statistical tests, then selects a specific number of features with the highest scores \cite{scikit-learn}. Various score-functions are available \textit{i.e.} ANOVA F-value, mutual information, chi-square, etc, each of which is suitable for a specific type of data. We selected ANOVA (analysis of variance) F-test to calculate the ratio between variances values and find a score for each feature saying how well this feature discriminates between three routes. Thereupon, remove the features with low scores that are independent of the routes.
\section{Routes Clustering:}
\label{Sec:Routes clustering}
Given a set of features for each climb, we can use a clustering model to make a certain number of clusters and group similar data from different climbs to the same cluster. 

K-Means clustering is a common algorithm and is widely used in literature. We set the number of clusters equal to three (equal to the number of routes), then the model initialized with three random center points in features space. Each data point (climb) is classified by computing the distance between that point and each group center. In other words, this model assigns cluster membership to each data point based on distance from cluster center \cite{scikit-learn}. Since cluster centers are randomly selected, this model does not give us a single answer. Hence, with only one model realization we can not have a definitive result and the algorithm must be verified through iterations. In the following for each set of features, K-Means clustering is evaluated over 100 iterations. 

Given the knowledge of the ground truth class assignments route-labels and our clustering algorithm assignments of the same samples predicted-labels, the (adjusted or unadjusted) rand-index is a function that measures the similarity of the two assignments, ignoring permutations. The average, maximum, and minimum value of rand-index over 100 iterations of K-Means clustering is shown by gray bars, blue lines, and solid lines respectively. Our reference for selecting the optimum number of features is based on a minimum number of features which leads to the highest value for the minimum rand-index. This way we consider the worst case, while the average and maximum rand-index could be much higher.
\begin{figure}[t!]
    \centering
    \vspace{-0.1cm}
    \includegraphics[width=0.65\textwidth]{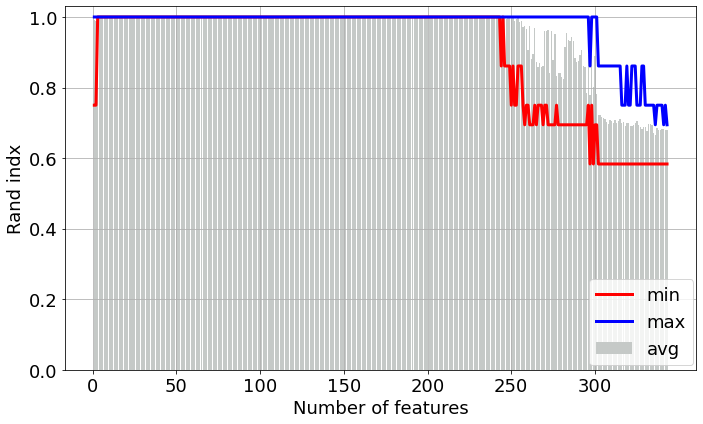}
    \vspace{-0.2cm}
    \caption{Feature optimization for K-Means clustering over $100$ iterations for data from climber 1 on $1^{st}$-day, features scaled by QuantileTransform and selected by ANOVA f-test}
    \vspace{-0.2cm}
    \label{fig:rand_index_climber1_day1}
    \end{figure}
\begin{itemize}
    \item {\bf Same climber in $1^{st}$ day (without prevision):} climber 1 on $1^{st}$-day.
    Figure \ref{fig:rand_index_climber1_day1} shows that with only three features, K-Means model clusters all climbs of climber 1 in $1^{st}$-day with rand-index = 1 in all 100 iterations. One feature is temporal and the other two are acceleration statistics. The temporal one is the time difference between clipping the $2^{nd}$ and the $5^{th}$ quickdraws: $\Delta t_{i_{k},j_{k}}^{c}$ where $i=5$, $j=2$, and $k=1$. The two statistical features belongs to the acceleration along y-axis of s-qd at position 7: 1- $(Y_{7}^{c})_{max}$ maximum value of acceleration, 2- $(Y_{7}^{c})_{95^{th}}$ the $95^{th}$ percentile of acceleration.
    Similar results were obtained for climber 2.
    \item {\bf Same climber in $2^{nd}$ day (with prevision):} climber 1 in $2^{nd}$-day. 
    \begin{figure}[t!]
    \centering
    \includegraphics[width=0.65\textwidth]{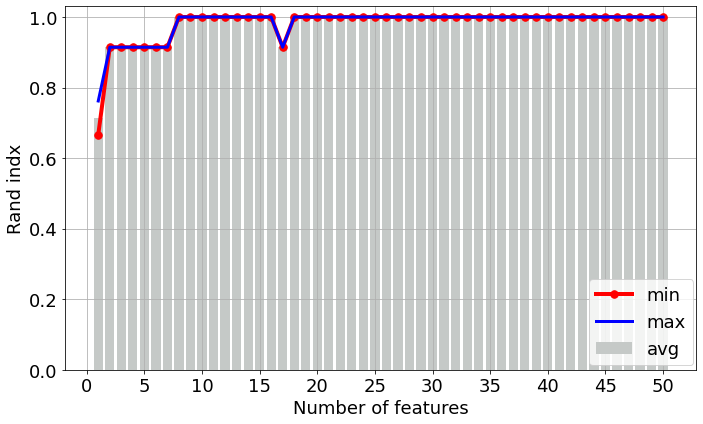}
    \vspace{-0.2cm}
    \caption{Feature optimization for K-Means clustering over $100$ iterations for data from climber 1 in $2^{nd}$-day, features scaled by QuantileTransform and selected by ANOVA f-test}
    \vspace{-0.2cm}
    \label{fig:rand_index_climber1_day2}
    \end{figure}
    Figure \ref{fig:rand_index_climber1_day2} demonstrates that despite the reordering of routes on the second day of the experiment (as described in subsection \ref{subsec:Contributions}), it is still possible to cluster the climbing data points for the same climber with a rand-index of 1. This clustering is achievable due to the climber's prior experience on these routes. However, factors like provisioning and fatigue alter the data in a way that, in contrast to Figure \ref{fig:rand_index_climber1_day1}, more features are necessary to achieve a rand-index of 1. These eight features are seven features from acceleration statistics of $5^{th}$ and $7^{th}$ s-qds and one temporal feature. The temporal feature is the average time spent between two subsequent s-qd given by $\overline{\Delta {ts}^{c}} = mean\left \{\Delta {ts}^{c} \right \}$. Hence, temporal features are not enough anymore and acceleration statistical features are of utmost importance. This implies that the selected statistical features belong to certain s-qds which are positioned in part of the line where different routes have different difficulties. As expected, the difficult part of the route is not at the beginning to avoid falling from a short distance to the ground and serious injuries.
    \item{\bf Same climber in different days:} 
    \begin{figure}[h!]
    \centering
    \vspace{-0.2cm}
    \includegraphics[width=0.65\textwidth]{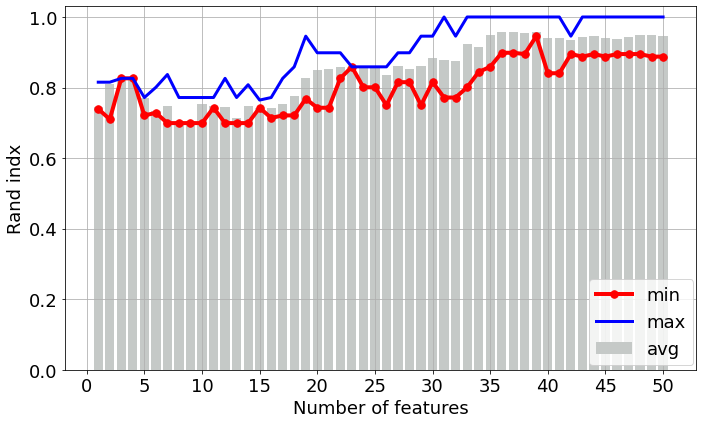}
    \vspace{-0.2cm}
    \caption{Feature optimization for K-Means clustering over $100$ iterations of mixing attempt 1 and 2 of climber 1, with features scaled by QuantileTransform and selected by ANOVA f-test}
    \vspace{-0.2cm}
    \label{fig:rand_index_climber1_day12}
    \end{figure}
    climber 1 on $1^{st}$ and $2^{nd}$ days.
    In this case, the best performance is rand-index equal to 0.94 obtained through 37 features. These features are extracted from s-qds at positions 2, 4, 5, and 7 and includes 35 acceleration statistical features and two temporal feature $\Delta t_{4_{1},2_{1}}^{c}$, $\Delta t_{5_{1},2_{1}}^{c}$. In Fig \ref{fig:rand_index_climber1_day12}, the first peak of minimum rand-index is 0.82 and obtained through 3 statistical features belonging to accelerations of s-qds at positions 5, and 7.
    \item {\bf Different climbers:} climber 1 on $1^{st}$ and $2^{nd}$ days and climber 2.
    The optimum performance in Fig \ref{fig:rand_index_climber12_day12} is obtained with 16 features: 14 acceleration statistics of s-qds at positions 5, 4, and 7 and two temporal features $\Delta t_{4_{1},2_{1}}^{c}$, $\Delta t_{5_{1},2_{1}}^{c}$.
    \begin{figure}[h!]
    \centering
    \vspace{-0.2cm}
    \includegraphics[width=0.65\textwidth]{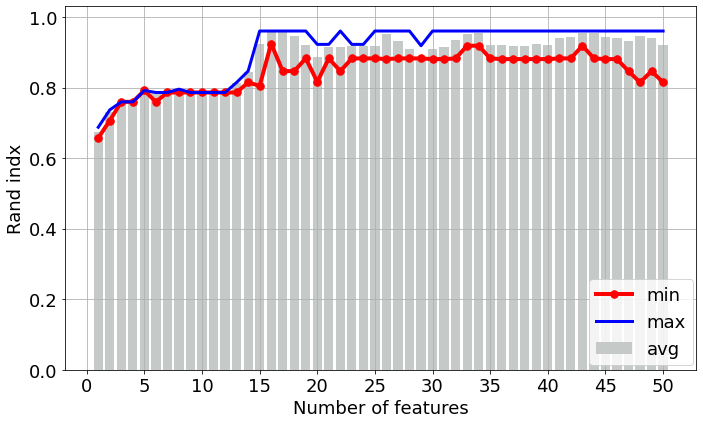}
    \vspace{-0.2cm}
    \caption{Feature optimization for K-Means clustering over $100$ iterations of mixing all attempts of two climbers, with features scaled by QuantileTransform and selected by ANOVA f-test}
    \vspace{-0.2cm}
    \label{fig:rand_index_climber12_day12}
    \end{figure}
\end{itemize}
\section{Visualization of clusters:}
\label{Sec:Visualization of Routes clustering}
\begin{figure}[t!]
\centering
\vspace{-0.1cm}
\includegraphics[width=0.65\textwidth]{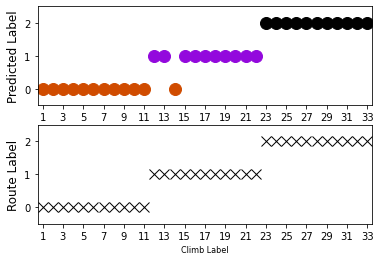}
\caption{K-Means clustering of climbing data points belongs to different climbers in different conditions}
\vspace{-0.3cm}
\label{fig:cluster123}
\end{figure}

Figure \ref{fig:cluster123} represents the outcome of clustering data points that correspond to climbing three different routes by different climbers in different conditions. For better visualization, the data points are sorted based on their labels from 0 to 2. Based on the results of the previous section, for each data point, 16 features are given to the K-Means model. As could be seen, only 1 climb (from route 1) out of 33 climbs has a predicted label (as for route 0) different from the other points belonging to the same route.

\begin{figure*}[ht]
\centering
\vspace{-0.1cm}
\includegraphics[width=0.9\textwidth]{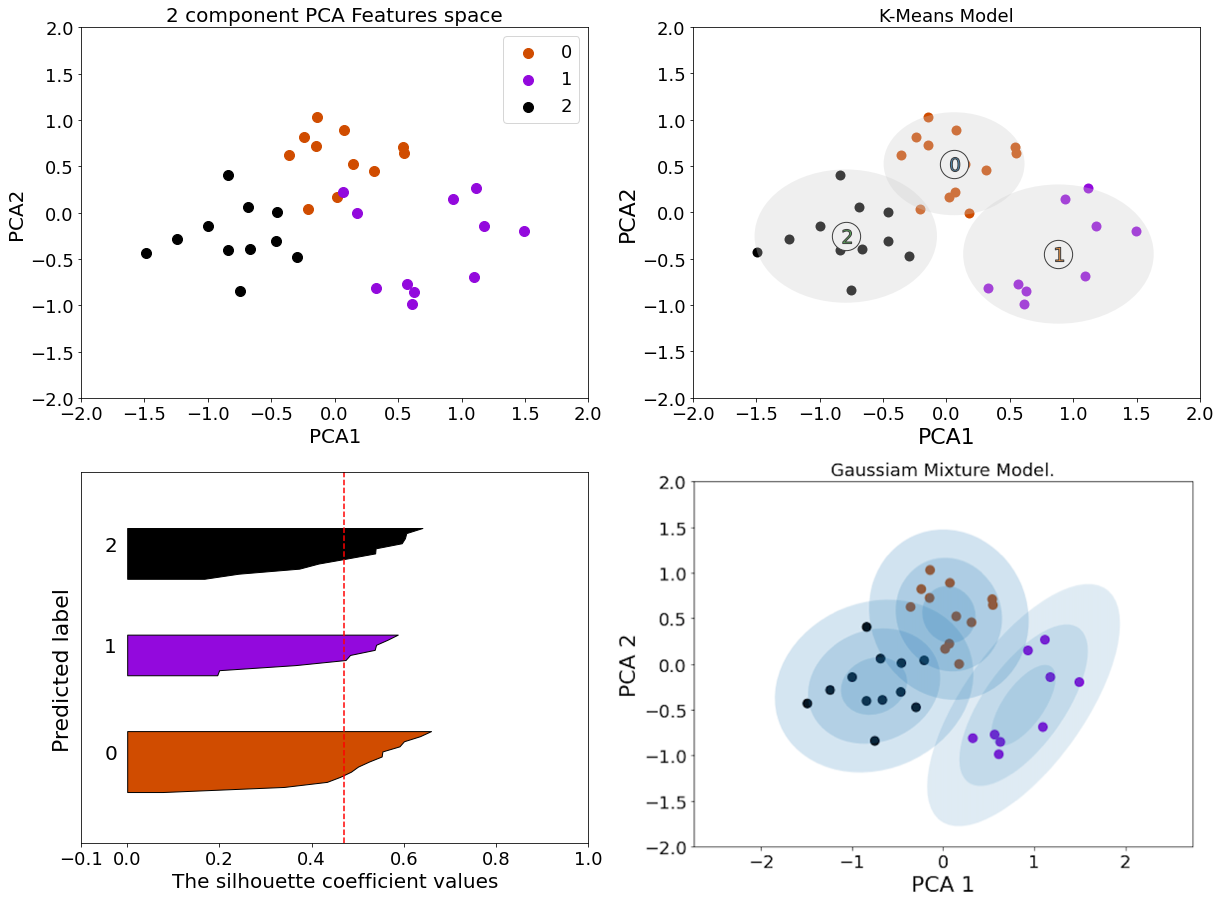}
\caption{Visualization of data points from different climbers in 2-D PCA feature space (top left), K-Means clustering (top right), silhouette scores for K-Means clusters (bottom left), and GMM clusterings (bottom right)}
\vspace{-0.3cm}
\label{fig:gmm_kmeans_pca_silh}
\end{figure*}
Figure \ref{fig:gmm_kmeans_pca_silh} depicts Principle Component Analysis (PCA) of data points in conjunction with K-Means and GMM clusterings. PCA is used to reduce the dimension of the selected features vector from 16 to 2. In 2-D PCA features space we can visualize how climbing data points are clustered by different models. However, for each climb, the K-Means model needs at least 5 features from PCA to have the optimum performance obtained with 16 selected features and without PCA. Here 2-D PCA slightly decreases the performance of the K-Means model and leads to one more mistake in data points clustering. Four subplots of Fig \ref{fig:gmm_kmeans_pca_silh} contains the following information:

\begin{itemize}
    \item {\bf top-left:} shows the ground truth in the 2-D features space (PCA1 and PCA2) as 33 data points refer to 33 climbs corresponding to 3 different routes identified by three colors (red, violet, black) and labels (0,1,2). 
    \item {\bf top-right:} we can see the three clusters' center points and the data points assigned to each of them. Despite the slight overlap between clusters 0 and 2, most of the data points are well-separated in the 2-D features space. Thus, the K-Means model finds suitable clustering results based on distance from centers. Here, with 2 features obtained from PCA, we have two data points from cluster 1 (violet) which are wrongly assigned to cluster 0 (red). 
    \item {\bf bottom-left:} The quality of the clusters is evaluated by silhouette score (on the left side) and shows how well climbing data points are grouped with other similar points within a cluster and separated from data points of other clusters. To do so, first, we compute the silhouette score for each data point. Then, sort the silhouette scores for data points belonging to each cluster. Consequently, for each cluster, we have a thick line with upper and lower limits along a vertical axis which refers to the maximum and minimum silhouette scores of its data points, respectively. The other samples fill the space between these two lines. It is evident that for each cluster, the thickness of different parts of the line depends on the number of points that have that specific silhouette score. The low value of the silhouette score means that the data point lies between two clusters. This is evident for cluster 0 and cluster 1 with silhouette scores less than 0.2 and 0.4, respectively which indicates their overlap in some of their samples.
    \item {\bf bottom-right:} In addition to the K-Means algorithm, the same data points are clustered through the Gaussian mixture model (GMM) and the result is depicted on the bottom right of Fig \ref{fig:gmm_kmeans_pca_silh}. Similar to K-Means, GMM clustering is also based on random initialization. But, clusters in GMM have flexible shapes based on a mixture of multi-dimensional Gaussian probability distributions that best fits all data points. While in K-Means each cluster is associated with a hard-edged sphere \cite{scikit-learn}. In this figure for each cluster, we have three ellipses of different sizes. The dimension of the smallest ellipses depends on the eigenvalues of the covariance matrix of the corresponding cluster and the other two ellipses are twice and three times bigger than the smallest one. The intensity of the color of each ellipse depends on the weight of the cluster evaluated by the GMM model. The higher the density of data points, the darker the ellipse. 
\end{itemize}

\section{Conclusion and Future work}
\label{Sec:Conclusion}
In this work, we utilized 3-axis accelerometer sensors working in ultra-low power mode and attached them to the quickdraws hanging from climbing walls. Features are extracted from sensors data which were measured while different climbers were climbing three different routes. Feature optimization reveals redundant features and corresponding s-qds, hence reducing the complexity of the algorithm and the hardware. The corresponding features are given to the K-Means algorithm to cluster data points to three different groups. 

The results show that when there is the possibility to identify the climber, we have perfect clustering with few features. However, the algorithm requires more features for detecting the routes climbed by the climber with prior practice, compared to the situation when the climber is without provisioning. 
But if data obtained from more than one climber and the climbers are not identified, then we need more features and the clustering algorithm may not be $100\%$ correct although the performance of the algorithm is still very high with only 1 mistake out of 33 climbs. In this case, the features are mainly statistical features obtained from the acceleration of s-qds which are positioned in the second half of the line. In addition to accelerations statistics, we need two temporal features as the time the climber spent to reach these s-qds. The position of these selected s-qds corresponds to the part of the line where different routes have different difficulties. Another reason why the statistical features obtained from the acceleration of lower sensors do not have any impact on clustering could be related to the behavior of the belayer. During each climb, the belayer pulled the rope tightly to prevent the climber from falling. Therefore, the s-qds at the beginning of the line can not move freely. 

As the next step, we aim first to collect more data from more climbers climbing other routes and improve the current algorithm. The second goal is to develop another algorithm based on a new type of feature which is the orientation of quickdraws. These features could be utilized to detect activities during climbing such as lowering. 
\section*{Acknowledgments}
This work has been partly supported by the project ``Sensors and data for the analysis of sport activities (SALSA)'', funded by the EFRE-FESR programme 2014-2020 (CUP: I56C19000110009). This  work  was supported  in  part  by  the  Italian  Ministry  of  Foreign  Affairs  and  International Cooperation, grant number US23GR04 (CUP: D43C23000350001).
\bibliographystyle{splncs04}
\bibliography{references}
\end{document}